\documentclass[12pt,preprint]{aastex} 

\shorttitle{ Neon Lights }
\shortauthors{Schmelz et al.}

\begin{document}

\title{Neon Lights Up a Controversy: the Solar Ne/O Abundance}

\author{J.T. Schmelz, K. Nasraoui, J.K. Roames, L.A. Lippner, J.W. Garst}
\affil{Physics Department, University of Memphis, Memphis, TN 38152}
\email{jschmelz@memphis.edu } 

\begin{abstract}
The standard solar model was so reliable that it could predict the existence of the massive neutrino. Helioseismology measurements were so precise that they could determine the depth of the convection zone. This agreement between theory and observation was the envy of all astrophysics -- until recently when sophisticated three-dimensional hydrodynamic calculations of the solar atmosphere reduced the metal content by a factor of almost two. Antia \& Basu (2005) suggested that a higher value of the solar neon abundance, $A_{Ne}/A_O=\ $0.52, would resolve this controversy. Drake \& Testa (2005) presented strong evidence in favor of this idea from a sample of 21 {\it Chandra} stars with enhanced values of the neon abundance, $A_{Ne}/A_O=\ $0.41. In this paper, we have analyzed solar active region spectra from the archive of the Flat Crystal Spectrometer on {\it Solar Maximum Mission}, a NASA mission from the 1980s, as well as full-Sun spectra from the pioneering days of X-ray astronomy in the 1960s. These data seem consistent with the standard neon-to-oxygen abundance value, $A_{Ne}/A_O=\ $0.15 (Grevesse \& Sauval 1998). If these results prove to be correct, than the enhanced-neon hypothesis will not resolve the current controversy.
\end{abstract}


\section{Introduction}

Scientific progress is made as theory evolves to predict observables and observations confirm theoretical predictions. There was no better astronomical example of this progress than the agreement between the complex models and helioseismology measurements of the solar interior -- that is, until recently. New 3-D hydrodynamic models (Allende Prieto, Lambert \& Asplund 2001; Asplund et al. 2000) which account for temperature inhomogeneities, convection and the associated velocity fields, and relax the assumption of local thermodynamic equilibrium have been used to compute atomic level populations. When these results are applied to solar absorption lines, the photospheric abundances of light elements such as C, N, O, and Ne all must be reduced by 25-35\% (Asplund et al. 2004). But these elements provide a major source of opacity for the solar interior, which determines the internal solar structure and the depth of the convection zone. The changes are significant enough to shatter the harmonious agreement between theory and observation that had come to exemplify scientific progress. The discrepancies between the two are now far in excess of either the observational uncertainties or the model predictions (Bahcall, Serenelli \& Pinsonneault 2004).

Antia \& Basu (2005) and Bacall, Basu \& Serenelli (2005) have suggested that one way to restore the disrupted harmony is to increase the solar abundance of neon. Where the C, N, and O abundances are determined from photospheric absorption lines, the neon abundance must be inferred from solar energetic particles and the spectra of hot stars and nebulae; it is often scaled to the oxygen abundance: $A_{Ne}/A_O=\ $0.15 (Grevesse \& Sauval 1998). If this value were 3.44 times higher, or $A_{Ne}/A_O=\ $0.52, the discord between the solar models and the helioseismology observations would disappear. 

To this end, Drake \& Testa (2005) determined the Ne/O abundance ratios for a sample of 21 stars located within 100 pc of the Sun which had been observed with the High Energy Transmission Grating on the {\it Chandra} X-ray Observatory. These results were possible because the X-ray spectra of cool stars (including the Sun) feature prominent emission lines of highly ionized ions of both oxygen and neon. These authors correctly point out that the abundances of neon and oxygen may be different in the photospheres and coronae of these stars -- they are both elements with high First Ionization Potential (FIP) -- but the {\it ratio} of the abundances should be identical. Their {\it Chandra} results suggest that $A_{Ne}/A_O=\ $0.41, significantly higher than the accepted solar value, and high enough (assuming other uncertainties) to resolve the controversies between the solar models and the helioseismology observations. In this paper, we have reexamined X-ray spectra similar to those of Drake \& Testa (2005), but for the Sun itself.

\section{Analysis}

The spectra analyzed here were obtained with the Flat Crystal Spectrometer (FCS -- Acton et al. 1980) on {\it Solar Maximum Mission}, which operated from 1980-89. The instrument had a 15 arcsec field-of-view and could scan the soft X-ray resonance lines of prominent ions in the range of 1.5~\AA\ to 20.0~\AA\ with a spectral resolution of 0.015 at 15 \AA. These lines are sensitive to a broad range of plasma temperatures (1.5--50 MK) and can be used to determine temperature, emission measure, density, and relative elemental abundances. The FCS observing sequence used to accumulate these spectra began with a low-resolution image of the active region. The instrument was then pointed at the brightest pixel of that image and began scanning at a rate of about 0.01 \AA\ per second from 13 \AA\ to 19 \AA. Shorter wavelengths were scanned simultaneously in different channels, but all the lines used in this Letter are in this limited range. Covering the full wavelength range took approximately 10 minutes, and four to five of these spectroscopic scans were done during the daylight portion of the 90-minute spacecraft orbit. 

These FCS spectra have been used in the past to study elemental abundances, but different science goals were driving those analyses: Strong, Lemen \& Linford (1991) searched for statistically significant abundance variations; Schmelz (1993) discovered abundance anomalies in flaring solar plasma; Fludra \& Schmelz (1995) found a set of absolute coronal abundances; and Schmelz et al. (1996) looked for deviations from the standard FIP effect. In none of these cases were the data used to compute the {\it normal} solar coronal Ne/O abundance ratio because, at the time, that value was "known." Now that doubt has been cast on the value of this ratio, we felt it was important to examine the FCS data archive to see if the data could confirm the higher Ne/O abundance ratios proposed by Antia \& Basu (2005) and observed by Drake \& Testa (2005).

The FCS archive contains 75 of the relevant spectral scans described above. Those taken during flares and long duration events were excluded from this sample to eliminate physical conditions which changed with time.  We also excluded those spectra where the hot Fe~XIX $\lambda$13.52 line (peak formation temperature Log T $=$ 6.8 MK) was visible. The presence of this line would indicate dynamic, rather than quiescent, plasma conditions. Of the remaining spectra we selected 20 with the best signal-to-noise ratio in order to be as confident as possible that the fluxes we have measured are dominated by the lines of interest. We will be using three Ne~IX lines, three O~VIII lines, and a Fe~XVIII-to-Fe~XVII line ratio as a temperature diagnostic. The wavelengths, peak formation temperatures, and spectroscopic notation for these transitions are listed in Table 1. The FCS data for these lines are listed in Table 2, which includes the date and time of the observation, the solar active region number, the electron temperature determined from the flux ratio of Fe~XVIII $\lambda$14.21 to Fe~XVII $\lambda$16.78, as well as the fluxes and uncertainties in photon units for the relevant neon and oxygen lines.

If we assume that the plasma in the FCS field-of-view is isothermal, then the observed flux, $F$, of a soft X-ray emission line is proportional to the elemental abundance, $A$, and the contribution function, $G(T)$. Then the Ne/O abundance ratio can be computed as follows:

\begin{equation}
{A_{Ne} \over A_{O} }\ =\ {F_{Ne} \over F_{O} }\  \times\  { G_{O}(T) \over G_{Ne} (T)}
\end{equation}

\noindent
The contribution functions were taken from version 4.02 of the CHIANTI Atomic Physics database (Dere et al. 1997; Young et al. 2003), and we have used the ionization fractions of Mazzotta et al. (1997). The $G(T)$ functions for the different neon and oxygen lines used in this analysis are plotted in the left column of Figure 1 and the $ G_{O}(T)$-to-$G_{Ne}(T)$ ratios for different pairs of the neon and oxygen lines are plotted in the right column. We can see from these plots that the ratios have a significant temperature dependence. Drake \& Testa (2005) added a fraction of the Ne~X~$\lambda$12.13 flux to the Ne~IX~$\lambda$13.45 flux to remove much of this temperature dependence: $ G_{Ne}\ =\ G_{Ne\ IX}\ +\ 0.15G_{Ne\ X}$. Unfortunately, however, this Ne~X $\lambda$12.13 line was not part of the FCS spectral scan. If we assume that the plasma in the FCS field-of-view is isothermal, we can compute the temperature from the flux ratio of Fe~XVIII $\lambda$14.21 to Fe~XVII $\lambda$16.78; these results are listed in Table 2. However, it is much more likely that the observed plasma is multi-thermal, with a Differential Emission Measure (DEM) typical of a quiescent active region (see, e.g., Brickhouse, Raymond \& Smith 1995; Brosius et al. 1996; Schmelz et al. 2001). If this is true, then the plasma temperatures listed in Table 2 are the mean values weighted by the DEM itself and the contribution functions of the two iron lines.

Unfortunately, it is not possible to construct a full DEM for the observed active region plasma using the FCS lines alone. (This is possible for a flare, however; see, e.g., Schmelz 1993). The problem is that the lowest temperature lines in the data set are the O~VIII lines listed in Table 1, with a peak formation temperature of Log T $=$ 6.5. Typical DEM curves of quiescent active region plasma have significant contributions at lower temperatures. So, although the high-temperature end of the DEM curve would be well-constrained by observations of the Fe~XVIII $\lambda$14.21 line flux and upper limits of the Fe~XIX $\lambda$13.52 line flux, the FCS lines, unfortunately, do not provide a useful low-temperature constraint.

We circumvent this problem in the following way: although it is not possible to determine the exact Ne/O abundance ratio without assuming the shape of the DEM curve, it {\it is} possible to determine an upper limit for this value. So, instead of searching for the precise Ne/O abundance ratio, we seek to determine if the FCS data are consistent with a Ne/O abundance ratio as high as $A_{Ne}/A_O=\ $0.52, the value proposed by Antia \& Basu (2005) which would eliminate the discord between the solar models and the helioseismology observations, or even as high as $A_{Ne}/A_O=\ $0.41, the value obtained by Drake \& Testa (2005) using the {\it Chandra} observations for a sample of 21 nearby stars.

To calculate this upper limit, we examine the plots in Figure 1.  In the temperature range of the typical quiescent active region DEM (6.0 $<$ Log T $<$ 6.6), the $ G_{O}(T)$-to-$G_{Ne}(T)$ ratios in Figure 1 all show a local maximum at Log T $\approx$ 6.3. Therefore, the upper limit of the Ne/O abundance ratio will result if we assume that the plasma temperature is Log T $\approx$ 6.3. This assumption is made in the analysis shown in Figure 2 where we have used Eq. 1 to calculate ${A_{Ne} / A_{O} }$ with ${F_{Ne} / F_{O} }$ from the FCS observations for the 20 spectra listed in Table 2, and ${ G_{O}(6.3) / G_{Ne} (6.3)}$ from the CHIANTI database. Each panel of Figure 2 also shows the weighted mean (solid line), the 1$\sigma$ uncertainty (dotted line), $A_{Ne}/A_O=\ $0.41 (dot-dash line) from Drake \& Testa (2005), and $A_{Ne}/A_O=\ $0.52 (dash line) from Antia \& Basu (2005). The panels of Figure 2 show the following ratios (a) Ne~$\lambda$13.45-to-O~$\lambda$16.01; (b) Ne~$\lambda$13.55-to-O~$\lambda$18.97; and (c) Ne~$\lambda$13.77-to-O~$\lambda$15.18. The weighted mean and its uncertainty are printed in the upper left corner of each plot. (Note: we show only three Ne/O ratios here, but all ratio possibilities gave similar abundance results.)

The abundances determined from the FCS fluxes and the CHIANTI atomic data seem to be in agreement with the standard solar value ($A_{Ne}/A_O=\ $0.15; Grevesse \& Sauval 1998), even though the results from Figure 2 represent the {\it highest} possible value for the Ne/O abundance ratio for these active region spectra. These results are not in agreement with the {\it Chandra} values observed in stars, and therefore, do not support the hypothesis put forth by Antia \& Basu (2005) -- that an enhanced abundance ratio ($A_{Ne}/A_O=\ $0.52) would eliminate the discord between the solar models and the helioseismology observations. In light of these results, we have also reconsider several observations of the Ne~IX and O~VIII spectral lines from the pioneering days of X-ray astronomy. These full-Sun instruments had the advantage of observing the Sun as a star, as {\it Chandra} does, and were therefore not weighted by the smaller FCS field of view.

(1) Blake et al. (1965) describe results from a U.S. Naval Research Laboratory Aerobee rocket flight on 1963 July 25. The payload included Bragg crystal spectrometers, pinhole cameras, and slit scanners. The spectral lines detected were attributed to a single active region near disk center. These lines included Ne~IX~$\lambda$13.44, O~VIII~$\lambda$16.01, and O~VIII~$\lambda$18.97 with fluxes of 2.1e4, 5.0e4, and 3.6e5 photons~cm$^{-2}~$s$^{-1}$, respectively. Using the CHIANTI atomic data and a temperature of Log T = 6.3, as we did for the FCS data, we find a maximum value for the neon-to-oxygen average abundance ratio of $A_{Ne}/A_O=\ $0.18.  

(2) Evans \& Pounds (1968) observed the full Sun with two slitless Bragg spectrometers flown on a Skylark rocket on 1966 May 5. There were two overlapping sets of spectra from two visible active regions with intensities in ergs~cm$^{-2}~$s$^{-1}$ of
Ne~IX~$\lambda$13.44 $=$ 18., 7.8; 
Ne~IX~$\lambda$13.70 $=$ 7.4, 3.9; 
O~VIII~$\lambda$15.18 $=$ 12., 7.; 
O~VIII~$\lambda$16.01 $=$ 20., 10.; and 
O~VIII~$\lambda$18.97 $=$ 130., 54. 
Averaging all the possible combinations gives $A_{Ne}/A_O=\ $0.25.

(3) In a series of papers, Walker, Rugge \& Weiss (1974) describe results from U.S. Air Force satellites OV1-10 from 1967 January 4 and OV1-17 from 1969 March 20. The OV1-10 Bragg crystal (flat KAP) spectrometer and a proportional counter spectrometer which viewed the Sun directly were mounted on a compact solar pointer. The instrument detected intensities in ergs~cm$^{-2}~$s$^{-1}$ for Ne~IX~$\lambda$13.44 $+$ $\lambda$13.55 $=$ 28.8; and O~VIII~$\lambda$18.97 $=$ 231. This gives $A_{Ne}/A_O=\ $0.20.

\section{Discussion}

The FCS data analyzed here and the rocket observations of the full X-ray Sun are consistent with the accepted solar abundance ratio of $A_{Ne}/A_O=\ $0.15. It is disappointing that these data do not confirm the prediction of Antia \& Basu (2005) of $A_{Ne}/A_O=\ $0.52, but it is not necessarily surprising. We have no doubt that the harmony between the solar models and the helioseismology observations will eventually be restored. What is somewhat surprising, however, is that this abundance result is not consistent with the value of $A_{Ne}/A_O=\ $0.41 obtained by Drake \& Testa (2005) for their sample of 21 nearby stars observed with {\it Chandra}. 

There are, however, a few published results where the solar Ne/O abundance seems to be statistically greater than the accepted value. Murphy et al. (1991) analyzed data from the {\it Solar Maximum Mission} Gamma-Ray Spectrometer for the X5 limb flare of 1981 April 27. They used a Bessel function for the accelerated particle spectrum and found $A_{Ne}/A_O>\ $0.35. Ramaty et al. (1996) used a power law rather than a Bessel function and revised the abundance to $A_{Ne}/A_O =\  $0.25. Schmelz (1993) studied FCS observations of the impulsive M5 flare on 1980 November 5 and found $A_{Ne}/A_O=\ $0.32.

Since the Ne/O abundance ratio is the same for all layers of the solar atmosphere, these enhanced neon detections were explained via a mechanism that would preferentially dredge up neon from the photosphere. Shemi (1991) suggested that preflare soft X-ray radiation could penetrate deep into the solar atmosphere and create nonthermal ionization ratios at the base of the chromosphere. Because the photoionization cross section of neon is higher than that of oxygen, ionized neon could be selected along with the thermally ionized low-FIP elements for preferential transfer to higher levels of the solar atmosphere by the ion-neutral differentiating mechanism operating in the chromosphere. Therefore, the layer where the energetic particles and the ambient plasma interact during the flare would contain an over-abundance of ionized neon with respect to other high-FIP elements like oxygen. The model requires an extended period of energetic soft X-ray emission to build up the supply of ionized neon. This condition was met for both neon-enhanced flares: the 1981 April 27 flare was a long duration event which lasted several hours in soft X-rays and the 1980 November 5 flare had a long and intense preflare phase.

Since the {\it Chandra}\ spectra analyzed by Drake \& Testa (2005) are most likely dominated by photons from very intense flares, it is possible that the enhanced neon abundances they measure result from a similar mechanism. Unfortunately, however, it does not seem likely that these higher neon abundances reflect the overall solar composition. If this result is true, than the proposed enhanced-neon hypothesis will not resolve the current controversy, and we will have to look elsewhere in order to restore the disrupted harmony between theory and observation.

We would like to thank E. DeLuca, S. Basu, J. Drake, P. Testa, N. Brickhouse, and M. Coffee for inspiring discussions on abundances. JTS would like to thank the Harvard-Smithsonian Center for Astrophysics for their kind hospitality during her 2004-05 sabbatical. Solar physics research at the University of Memphis is supported by NSF ATM-0402729 and NASA NNG05GE68G. 

{}

\begin{figure}
\figurenum{1}
\epsscale{.9}
\plotone{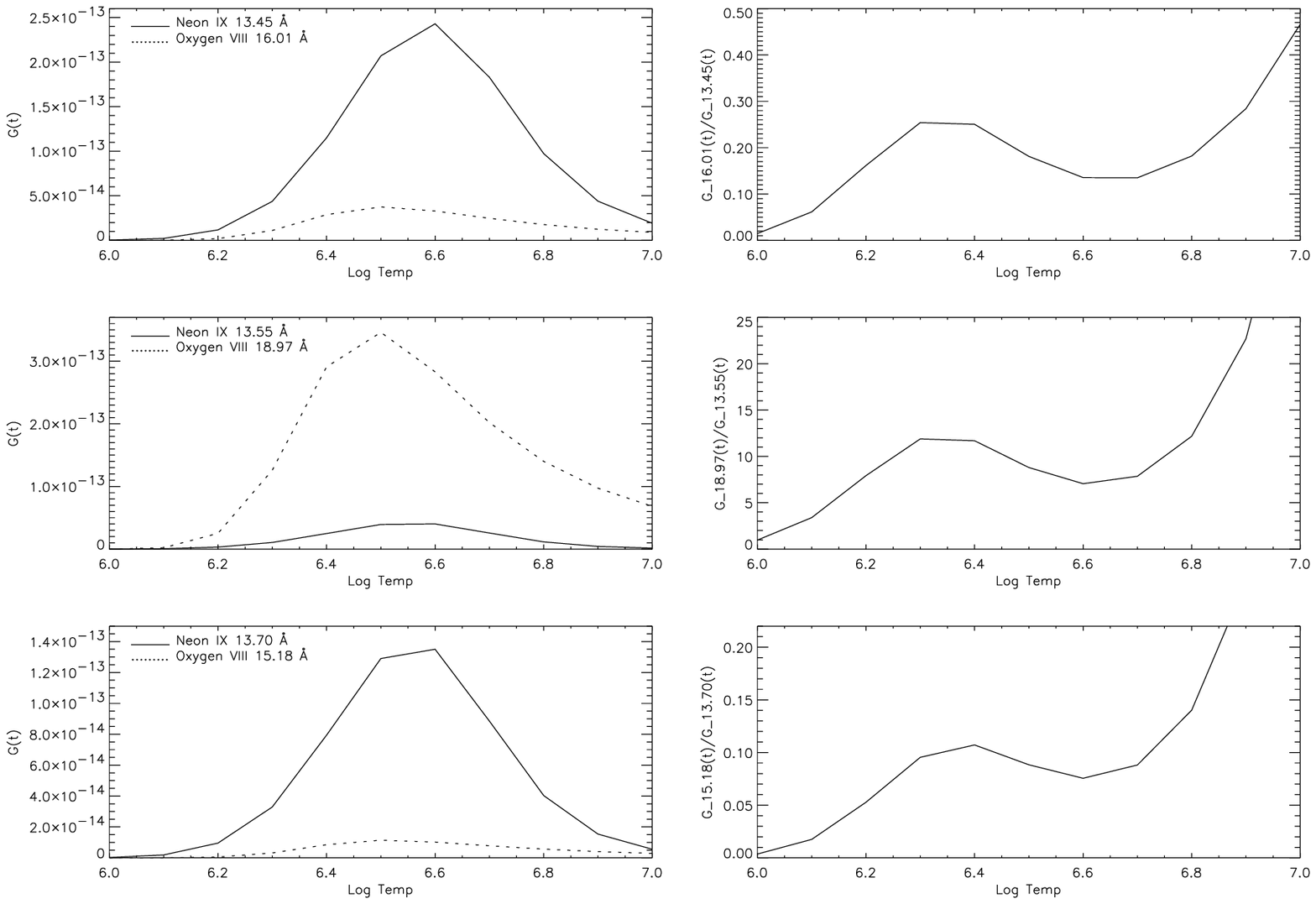}
\caption{G(T) functions from CHIANTI for pairs of neon and oxygen lines (left) and ratios of these pairs (right). 
(a) Ne~IX~$\lambda$13.45-to-O~VIII~$\lambda$16.01; 
(b) Ne~IX~$\lambda$13.55-to-O~VIII~$\lambda$18.97; and 
(c) Ne~IX~$\lambda$13.77-to-O~VIII~$\lambda$15.18. 
Each of the ratios has a local maximum at Log T $=$ 6.3, the temperature used in Figure 2 to determine the maximum Ne/O abundance ratio consistent with the FCS active region spectra.
}
\end{figure}

\begin{figure}
\figurenum{2}
\epsscale{.9}
\plotone{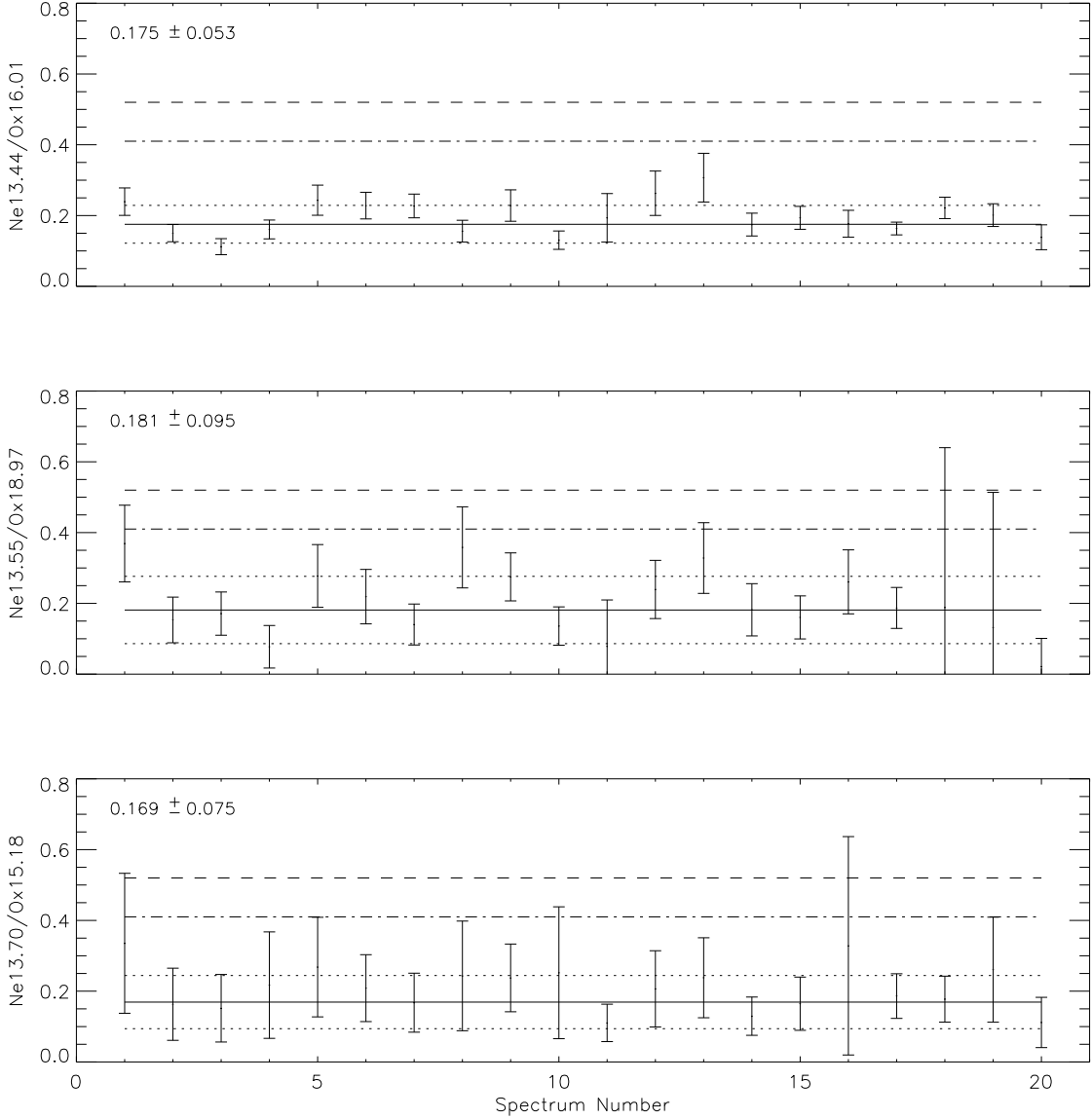}
\caption{${A_{Ne} / A_{O} }$ with ${F_{Ne} / F_{O} }$ from the FCS observations for the 20 spectra listed in Table 2, and ${ G_{O}(6.3) / G_{Ne} (6.3)}$ from the CHIANTI database: (a) Ne~IX~$\lambda$13.45-to-O~VIII~$\lambda$16.01; 
(b) Ne~IX~$\lambda$13.55-to-O~VIII~$\lambda$18.97; and 
(c) Ne~IX~$\lambda$13.77-to-O~VIII~$\lambda$15.18.
The solid line is the weighted mean, dotted lines are $\pm$1$\sigma$ uncertainties, dot-dash line is $A_{Ne}/A_O=\ $0.41 (Drake \& Testa 2005), and the dash line is $A_{Ne}/A_O=\ $0.52 (Antia \& Basu 2005). The weighted mean and its uncertainty are printed in the upper left corner of each plot.
}
\end{figure}

\begin{deluxetable}{lllc}
\tabletypesize{\scriptsize}
\tablewidth{0pt}
\tablehead{
\colhead{Ion} & \colhead{$\lambda$ (\AA)} & \colhead{Log T} & \colhead{Transition} 
}
\startdata

Ne IX		& 13.45	& 6.6	& 1s$^2$\ $^1$S$_0$\ -- 1s2s\ $^1$P$_1$\\
Ne IX		& 13.55	& 6.6	& 1s$^2$\ $^1$S$_0$\ -- 1s2p\ $^3$P$_1$\\
Ne IX		& 13.70	& 6.6	& 1s$^2$\ $^1$S$_0$\ -- 1s2s\ $^3$S$_1$\\
O VIII		& 15.18	& 6.5	& 1s\ $^2$S$_{1/2}$\ -- 4p\ $^2$P$_{3/2,1/2}$\\
O VIII		& 16.00	& 6.5	& 1s\ $^2$S$_{1/2}$\ -- 3p\ $^2$P$_{3/2,1/2}$\\
O VIII		& 18.97	& 6.5	& 1s\ $^2$S$_{1/2}$\ -- 2p\ $^2$P$_{3/2,1/2}$\\
Fe XVIII	& 14.20	& 6.8	& 2s$^2$2p$^5$\ $^2$P$_{3/2}$\ -- 2p$^4$(1d) 3d\ $^2$D$_{5/2,3/2}$\\
Fe XVII	& 16.78	& 6.7	& 2p$^6$\ $^1$S$_0$\ -- 2p$^5$ 3s\ $^3$P$_1$\\

\enddata
\end{deluxetable}

\begin{deluxetable}{lllllllllll}
\tabletypesize{\scriptsize}
\tablewidth{0pt}
\tablehead{
\colhead{} & \colhead{Date} & \colhead{Time (UT)} & \colhead{AR} & \colhead{T(MK)} & \colhead{ Ne$\lambda$13.44} & \colhead{ Ne$\lambda$13.55} & \colhead{ Ne$\lambda$13.70} & \colhead{ O$\lambda$15.18} & \colhead{ O$\lambda$16.01} & \colhead{ O$\lambda$18.97} 
}
\startdata

1&1986 May 20 & 00:19-01:12 & 4729 &3.09$^{.26}_{.34}$ & 4300$.\pm$320. & 1100$.\pm$300. & 3300$.\pm$300. & 940$.\pm$550. & 4600$.\pm$660. & 35100$.\pm$3800.\\ 
2&1986 May 21 & 14:02-14:55 & 4731 &3.51$^{.12}_{.12}$ & 3300$.\pm$300. & 1360$.\pm$260. & 2100$.\pm$270. & 1200$.\pm$740. & 5600$.\pm$750. & 49700$.\pm$4000.\\
3&1986 May 23 & 02:11-03:07 & 4731 &3.35$^{.16}_{.19}$ & 1850$.\pm$240. & 690$.\pm$240. & 1760$.\pm$240. & 1100$.\pm$680. & 4200$.\pm$650. & 47600$.\pm$4300.\\ 
4&1986 May 24 & 04:54-05:50 & 4731 &3.47$^{.12}_{.19}$ & 3100$.\pm$300. & 300$.\pm$230. & 2400$.\pm$280. & 1100$.\pm$720. & 4900$.\pm$650. & 46000$.\pm$5400.\\ 
5&1987 Apr 11 & 22:26-23:21 & 4787 &3.59$^{.17}_{.16}$ & 5300$.\pm$400. & 1200$.\pm$370. & 4400$.\pm$380. & 1600$.\pm$800. & 5500$.\pm$870. & 52400$.\pm$5300.\\ 
6&1987 Apr 13 & 01:08-02:03 & 4787 &3.80$^{.13}_{.13}$ & 4300$.\pm$360. & 1000$.\pm$330. & 3700$.\pm$350. & 1700$.\pm$750. & 4800$.\pm$680. & 53200$.\pm$4700.\\ 
7&1987 Apr 14 & 14:50-15:45 & 4787 &3.76$^{.09}_{.13}$ & 4700$.\pm$340. & 700$.\pm$300. & 3000$.\pm$370. & 1700$.\pm$800. & 5200$.\pm$670. & 61200$.\pm$4700.\\ 
8&1987 Apr 15 & 23:22-24:17 & 4787 &3.76$^{.13}_{.13}$ & 7000$.\pm$690. & 1900$.\pm$600. & 5200$.\pm$600. & 2000$.\pm$1300. & 11300$.\pm$1900. & 64000$.\pm$9300.\\
9&1987 Apr 19 & 14:41-15:10 & 4787 &3.85$^{.09}_{.13}$ & 11000$.\pm$900. & 3200$.\pm$800. & 9400$.\pm$800. & 3800$.\pm$1500. & 12000$.\pm$2200. & 138000$.\pm$10200.\\
10&1987 Nov 27 & 16:20-17:17 & 4891 &3.27$^{.15}_{.18}$ & 3100$.\pm$330. & 680$.\pm$260. & 2800$.\pm$300. & 1100$.\pm$780. & 6100$.\pm$1000. & 59000$.\pm$4400.\\
11&1987 Nov 29 & 20:07-21:06 & 4891 &3.47$^{.12}_{.12}$ & 2400$.\pm$560. & 200$.\pm$73. & 2700$.\pm$600. & 2400$.\pm$1000. & 3100$.\pm$800. & 32300$.\pm$7500.\\
12&1987 Dec 07 & 14:48-16:01 & 4901 &3.63$^{.17}_{.20}$ & 3500$.\pm$300. & 720$.\pm$230. & 2200$.\pm$290. & 1000$.\pm$500. & 3400$.\pm$750. & 35600$.\pm$4300.\\
13&1987 Dec 08 & 09:49-10:51 & 4901 &4.03$^{.14}_{.18}$ & 5100$.\pm$400. & 1100$.\pm$320. & 3400$.\pm$350. & 1300$.\pm$620. & 4200$.\pm$890. & 40700$.\pm$4300.\\
14&1987 Dec 09 & 03:06-04:08 & 4901 &3.94$^{.09}_{.13}$ & 3500$.\pm$370. & 740$.\pm$300. & 2200$.\pm$320. & 1600$.\pm$640. & 5100$.\pm$770. & 48500$.\pm$4500.\\
15&1987 Dec 11 & 02:13-03:13 & 4901 &3.80$^{.09}_{.09}$ & 4900$.\pm$350. & 820$.\pm$300. & 3000$.\pm$310. & 1700$.\pm$780. & 6400$.\pm$970. & 60600$.\pm$4500.\\
16&1987 Dec 11 & 10:04-11:04 & 4901 &4.12$^{.10}_{.09}$ & 3700$.\pm$360. & 900$.\pm$300. & 2500$.\pm$320. & 740$.\pm$690. & 5300$.\pm$1000. & 41500$.\pm$4200.\\
17&1987 Dec 13 & 09:10-10:09 & SELimb &3.31$^{.16}_{.15}$ & 6400$.\pm$440. & 1400$.\pm$410. & 6100$.\pm$430. & 3100$.\pm$1030. & 10000$.\pm$850. & 86200$.\pm$5600.\\
18&1987 Dec 15 & 03:34-04:31 & 4906 &3.72$^{.13}_{.17}$ & 5900$.\pm$420. & 1100$.\pm$380. & 4200$.\pm$380. & 2300$.\pm$800. & 6800$.\pm$790. & 71000$.\pm$5900.\\
19&1987 Dec 16 & 07:49-08:46 & 4906 &3.80$^{.13}_{.13}$ & 4300$.\pm$360. & 620$.\pm$200. & 3600$.\pm$420. & 1300$.\pm$730. & 5400$.\pm$740. & 55800$.\pm$4800.\\
20&1987 Dec 18 & 03:46-04:42 & 4906 &2.95$^{.28}_{.96}$ & 1700$.\pm$270. & 56$.\pm$19. & 1000$.\pm$230. & 870$.\pm$520. & 3100$.\pm$600. & 30600$.\pm$4000.\\
\enddata
\end{deluxetable}

\end{document}